\documentstyle[aps,preprint]{revtex}

\begin{document}

\draft

\preprint{CECS 98, hep-th/9803034}

\title{Supersymmetric Particle in a Spacetime with Torsion and the Index Theorem}

\author{Osvaldo Chand\'{\i}a$^{a}${\thanks{E-mail: ochandia@cecs.cl}} and Jorge Zanelli$^{a,b}${\thanks{E-mail: jz@cecs.cl}}{\thanks{John Simon Guggenheim fellow.}}}

\address{$^{(a)}$ Centro de Estudios Cient\'{\i}ficos de Santiago,
Casilla 16443, Santiago, Chile\\
$^{(b)}$ Departamento de F\'{\i}sica, Universidad
de Santiago de Chile, Casilla 307, Santiago 2, Chile.}

\maketitle

\begin{abstract}

The supersymmetric Lagrangian compatible with the presence of torsion in the background spacetime requires, in addition to the minimal coupling, an interaction between the spin and the torsion of the form $\frac{1}{2} N_{\mu\nu\lambda\rho}\psi^\mu\psi^\nu\psi^\lambda\psi^\rho$, where $N$ is the Nieh-Yan 4-form. This gives rise to a coupling between helicity ($h$) and the Nieh-Yan density of the form $hN$.  The classical Lagrangian allows computing the index for the Dirac operator on a four-dimensional compact manifold with curvature and torsion using the path integral representation for the index.  This calculation provides an independent check for the recent result of the chiral anomaly in spaces with torsion. 

\end{abstract}

\pacs{PACS numbers: 02.40.Vh, 04.90.+e, 11.30.Pb, 12.60.Jv}


\section{Introduction}

Supersymmetry is not only an important property for constructing quantum field theories, but is also a useful guideline for describing the dynamics of spin degrees of freedom in the ``classical'' limit \cite{Casalbuoni,Galvao}, and for predicting the form of the interaction between the spin and external fields. Thus, the nonrelativistic spin-orbit coupling in a central potential, as well as the interaction between spin and various gauge and gravitational fields can be obtained \cite{GZ1,GZ2}.  

Out of the four fundamental interactions of nature, only electromagnetism and gravitation manifest themselves over regions much larger than the Compton wavelength of a typical particle. Although the interaction between spin and electromagnetism has been known since the discovery of spin itself, the interaction with gravitation has remained somewhat obscure.  In particular, the effect of torsion on a particle with spin is not completely understood: the Hamiltonian analysis indicates that contributions of the form $~\alpha \psi^{\mu}\psi^{\nu} T^{\lambda}_{\mu \nu}\hat{p}_{\lambda} + \beta \psi^{\mu} \psi^{\nu} \psi^{\lambda} \psi^{\rho} R_{\mu \nu \lambda \rho}$ should be present \cite{Galvao,GZ2}, but to our knowledge, the geometric significance of these terms had not been fully spelled out. 

\section{Supersymmetric spinning particle}

Here we construct the Lagrangian for the supersymmetric spinning particle in a torsional background, based on the following assumptions: i) the ordinary derivatives of the Bosonic coordinates are replaced by covariant ones, and ii) the supersymmetry algebra has the same form as the one in a flat background, namely, 

\begin{eqnarray} \label{delta}
\delta q^\mu &=& i\varepsilon \psi ^\mu , \nonumber \\
\delta \psi ^\mu &=& \varepsilon {\dot q} ^\mu .
\end{eqnarray}

Let us first review supersymmetry on flat spacetimes and on curved torsion-free spacetimes.  Consider the free spinning particle whose Lagrangian is given by

\begin{equation} \label{L0}
L _0= \frac{1}{2}\delta _{\mu\nu}{\dot q}^{\mu}{\dot q}^{\nu} + \frac{i}{2}\delta _{\mu\nu} \psi ^\mu {\dot \psi} ^\nu .
\end{equation}

This system is supersymmetric under (\ref{delta}) where $\varepsilon$ is a constant real Fermionic parameter. These transformations are generated by the supercharge  

\begin{equation} \label{Q0}
Q = \psi^\mu p_\mu ,
\end{equation}
using the (Dirac) brackets of the basic variables \cite{Galvao}, $\{ q^\mu , p_\nu \} = \delta ^\mu _\nu ,\;\; \{ \psi ^\mu , \psi ^\nu \} = i\delta ^{\mu\nu}$, and the definition $p_{\mu} = \frac{\partial L}{\partial {\dot q}^{\mu}} ={\dot q}_{\mu}$ has been used. The resulting supersymmetry algebra is $\{Q,Q\} = 2iH$, $\{H,Q\}=0$, where $H=\frac{1}{2} p^2$ is the Hamiltonian.

The system is quantized through the usual replacement of phase space variables $q$, $p$, $\psi$ by operators and symplectic structure by the standard forms
($\hbar$ is set equal to 1),
\begin{eqnarray} \label{[]}
[ {\hat q}^\mu , {\hat p}_\nu ] &=& i\delta ^\mu _\nu ,\nonumber \\
\{ {\hat \psi}^\mu , {\hat \psi}^\nu \} &=& \delta ^{\mu\nu}.
\end{eqnarray}
These relations are realized by the usual Schrodinger representation, ${\hat q}^\mu = q^\mu$, ${\hat p}^\mu = -i\partial _\mu$, while the Fermionic operators satisfy the Clifford algebra and therefore one identifies them with Dirac gamma matrices through ${\hat \psi}^\mu = \frac{1}{\sqrt{2}} \gamma^\mu$.

Thus, the quantum supersymmetric charge for the free superparticle is proportional to the Dirac operator

\begin{equation} \label{hatQ}
{\hat Q} = {\hat \psi}^\mu {\hat p}_\mu = \frac{i}{\sqrt{2}}\gamma^\mu\partial_\mu,
\end{equation}
and the quantum supersymmetry algebra replicates the classical one.

In the presence of an external torsion-free gravitational field the Lagrangian for a superparticle is \cite{Windey,Alv-G,Friedan,Barducci}, 
\begin{equation} \label{Lg}
L_g = \frac{1}{2}g_{\mu\nu}{\dot q}^\mu{\dot q}^\nu + \frac{i}{2}\psi_a ({\dot \psi}^a +{\stackrel{\circ}{\omega^a}}_{b\rho}\psi^b{\dot q}^\rho) ,
\end{equation}
where ${\stackrel{\circ}{\omega}}^{ac}_{\mu}=e^a_{\lambda}(e^{\lambda c},_{\mu}+ \stackrel{\circ}{\Gamma^\lambda}_{\mu\nu}e^{c\nu})$ is the metric spin connection. The canonical variables now are $q^{\mu}$ and $\psi^a$; the projected components of the spinor on the manifold --$\psi^{\mu}$-- are the composite variables $e_a^{\mu}\psi^a$, where $e_a^{\mu}$ is the inverse vielbein. Note that this Lagrangian changes by a total derivative under the original supersymmetry transformations (\ref{delta}) in terms of $q^{\mu}$ and $\psi^{\mu}$. The quantum supersymmetry generator is the correct Dirac operator, given by 

\begin{equation} \label{Q}
\hat{Q} = \frac{i}{\sqrt{2}}\gamma^\mu (\partial_\mu + \frac{1}{2}{\omega ^{ab}}_\mu J_{ab}) ,
\end{equation}
where $J_{ab} = \frac{1}{4}[\gamma_a,\gamma_b]$ is the generator of Lorentz transformations in the spinorial representation. 

In order to extend the Lagrangian (\ref{Lg}) to the torsionful case it is not sufficient to replace $\stackrel{\circ}{\omega}$ by the full spin connection, 

\begin{equation} \label{omega}
 {\stackrel{\circ}{\omega^a}}_{b\mu}\longrightarrow \omega^a_{b \mu} = {\stackrel{\circ}{\omega^a}}_{b\mu} + \kappa^a_{b \mu},
\end{equation}
where $\kappa^a_{b \mu} = e^{a\lambda} e_b^{\nu} T_{[\lambda \nu \mu]}$ is the contorsion tensor.  Indeed, the Lagrangian invariant under  (\ref{delta}) in this case reads  

\begin{equation} \label{L}
L = L_g -\frac{i}{2}{\dot q}^\nu \kappa_{\nu \lambda \rho} \psi ^\lambda \psi ^\rho - \frac{1}{8}N_{\mu \nu \lambda \rho} \psi ^\mu \psi ^\nu \psi ^\lambda \psi ^\rho
\end{equation}
where
\begin{equation} \label{Ncomp}
N_{\mu \nu \lambda \rho} =  g_{\alpha \beta} {T^\alpha}_{\mu \nu} {T^\beta}_{\lambda \rho} - 2R_{\mu \nu \lambda \rho},
\end{equation}
are the components of the Nieh-Yan tensor \cite{NY}, where $T^{\alpha}_{\mu \nu} = \Gamma^{\alpha}_{[\mu\nu]}$. The second term on the r.h.s. of (\ref{L}) results from the minimal substitution (\ref{omega}), while the third term is a nonminimal one which is required by supersymmetry alone. In the first-quantized form of this system in the Hamiltonian formalism this term {\em{does}} arise by minimal substitution through the expression $\frac{i}{2} \sigma^{ab}[e_a^{\mu}\nabla_{\mu}, e_b^{\nu}\nabla_{\nu}]$, where $\nabla_{\mu}$ is the full covariant derivative (see, Ref.\cite{GZ2}).

\section{Quantum Mechanics}

In the first-quantized theory the relevant operator to describe spinning states is the Dirac operator, which in a generic spacetime is obtained via the minimal substitution $\partial_{\mu} \rightarrow \partial_{\mu} + \omega_{ab\mu}J^{ab}$,

\begin{equation} \label{D}
D = i(\gamma^a e_a^{\mu} \partial_{\mu} + \gamma^{ab} \gamma^c t_{abc}) ,
\end{equation}
where $t_{abc}= \omega_{ab\mu} e_c^{\mu}$, and $\gamma^{ab} = \frac{1}{4}[\gamma^a,\gamma^b]$ is the generator of Lorentz transformations in the spinorial representation.  

A massless Fermi field satisfying $D \Psi =0$ is invariant under global chiral transformations and has an associated conserved current $J^\mu_5(x)$. At the second-quantized level this chiral symmetry is broken, which is reflected by the nonconservation of the expectation value of the chiral current      

\begin{equation} \label{J}
\partial_\mu \left\langle \hat{J}^\mu_5(x) \right\rangle = {\cal A}(x) .
\end{equation}

In four spacetime dimensions the anomaly ${\cal A}$ is given by the second Chern character,  also known as the Pontryagin density \cite{Salam}, for the rotation group on the tangent space  plus a torsional piece \cite{CZ}, 

\begin{equation} \label{chandia}
{\cal A}(x) = \frac{1}{8\pi^2} [R^{ab} \mbox{\tiny $\wedge$} R_{ab} + 2(T^a \mbox{\tiny $\wedge$} T_a - R_{ab} \mbox{\tiny $\wedge$} e^a \mbox{\tiny $\wedge$} e^b)] ,
\end{equation}
where ${R^a}_b = d{\omega^a}_b + {\omega^a}_c \mbox{\tiny $\wedge$}  {\omega^c}_b$ (Riemann curvature), and $T^a = de^a + {\omega^a}_b \mbox{\tiny $\wedge$} e^b$ (torsion two-form). The last terms on the right hand side of (\ref{chandia}) are the Nieh-Yan four-form $N=N_{\mu  \nu \lambda \rho} dx^{\mu} \mbox{\tiny $\wedge$} dx^{\nu} \mbox{\tiny $\wedge$} dx^{\lambda} \mbox{\tiny $\wedge$} dx^{\rho}$, which  vanish identically if torsion tensor is zero. 

In this calculation, the tetrad one-form $e^a$ is defined in {\em geometrical} units, that is to say, units in which all differential forms are dimensionless.  In these units, the Dirac operator (\ref{D}) is dimensionless as well and the tetrad one form $e^a$ can now be viewed as part of a connection on equal footing with $\omega^{ab}$.  This allows for an embedding of the local $SO(4)$ group into $SO(5)$ as 

\begin{equation} \label{38}
W^{AB} = \left(
\begin{array}{cc}
\omega^{ab} & \,\,e^a\\
- e^b & \,\,0
\end{array} \right),
\end{equation}
where $a,b = 1,\cdots 4$ and $A, B =1,\cdots 5$.  In this way, as was shown in \cite{CZ}, the integral of (\ref{chandia}) is the Chern character for $SO(5)$. Evaluating the integral on a compact four-dimensional Euclidean manifold yields an integer because $\pi_3 [SO(5)] = {\bf Z}$ \cite{Nakahara}.

This result may seem paradoxical: If the Dirac operator is obtained purely by the minimal substitution mentioned above, the standard argument would lead to believe that the Chern character relevant for the anomaly is the one constructed from the curvature for spin connection, namely the Pontryagin form $\frac{1}{8\pi^2}R_{ab}\mbox{\tiny $\wedge$} R^{ab}$. However this is incorrect: the extra {\em{nonminimal}} coupling to the vielbein in (\ref{D}) brings in torsion through the commutator
\begin{equation}\label{com}
[\nabla_{\mu}, \nabla_{\nu}] = \frac{1}{2} R^{ab}_{\mu \nu}J_{ab} - T^{\lambda}_{\mu \nu} \nabla_{\lambda}. 
\end{equation}

A related paradox comes from writing the (Hermitian) Dirac operator as $\gamma^{\mu} {\nabla}_{\mu}$, where  $\gamma^{\mu}:=\gamma^a e_a^{\mu}$, $D_{\mu}= \partial_{\mu} + t_{\mu}$, and $t_{\mu} = \frac{1}{3!}t_{abc}\epsilon^{abcd}e_{d \mu}$ is the dual of the ``H-torsion''. This approach could lead to erroneously interpret the effect of torsion as equivalent to the addition of a U(1)  connection. The point is that $\gamma^{\mu}= \gamma^a e_a^{\mu}$ is not covariantly constant.

\section{Index Theorem}

We can now show that the nonminimal coupling in the classical Lagrangian is precisely responsible for the anomaly in the second-quantized theory. This can be shown using the path integral for a supersymmetric particle as a representation the index for the Dirac operator.  This is an independent calculation of the anomaly for spaces with torsion obtained in \cite{CZ}, which didn't resort to supersymmetry. 

The integral of the chiral anomaly is the index of the Dirac operator under $\gamma_5$-conjugation \cite{Fujikawa}. As seen above, the Dirac operator is the generator of supersymmetry in the first-quantized form. Hence, the anomaly is just the index of $\hat{Q}$,
\begin{equation}\label{ind}
\int {\cal A}(x) = \mbox{ind}\hat{Q},
\end{equation}
which is also the Witten index, Tr$[(-1)^F]$ \cite{Windey}. As shown in \cite{Cecotti} the regularized Witten index admits a path integral representation 

\begin{equation} \label{path}
\mbox {Tr} [(-1)^F \exp{(-\hat{H})}] = \int_{\mbox{\tiny{PBC}}} Dq D\psi e^{-\int_0^1 L( q,\dot{q},\psi,\dot{\psi} )},
\end{equation}
The different ingredients in (\ref{path}) are the following:  $\hat{H}$ is the quantum Hamiltonian whose classical Lagrangian is $L$, and is the square of the supersymmetry generator, $\{\hat{Q},\hat{Q}\}= 2\hat{H}$; the conjugation operator $(-1)^F$ anticommutes with $\hat{Q}$; PBC stands for periodic boundary conditions in Euclidean time for all fields ($q(0) = q(1)$, etc.); finally, the index is ind$\hat{Q} = n^{(0)}_+ - n^{(0)}_-$, where $n^{(0)}_{+ (-)}$ is the number of Bosonic (Fermionic) zero-energy states.

The path integral is evaluated in the stationary phase approximation taking $q^\mu = \bar{q}^\mu + \xi^\mu$, where $\bar{q}$ is the classical (periodic) trajectory and $\xi^\mu$ is a small fluctuation around it (periodic as well). In this approximation the classical trajectories are assumed to have vanishing fermionic fields.  The terms quadratic in $\xi$ in the action give the dominant contribution to the index.  In a generic curved background the coordinates can always be chosen so that the connection vanishes on a line (normal coordinates).  Choosing normal coordinates along the entire classical path, the action is, up to terms quadratic in $\xi$ reads

\begin{eqnarray} \label{action}
S &=& S_0[{\bar q} , \psi] + \int_0^1 \frac{1}{2} \eta_{\mu \nu}\dot{\xi}^\mu \dot{\xi}^\nu +\nonumber \\
& &\frac{i}{4} \xi^\lambda {\dot \xi}^\rho (R_{\mu\nu\lambda\rho}  +T_{\lambda\mu\nu,\rho} -  T_{\rho\mu\nu,\lambda}) \psi^\mu \psi^\nu .
\end{eqnarray}

Since the variables are periodic in time, they can be expanded in Fourier modes ${\xi_n^{\mu}}$, ${\psi_n^\mu}$ (with the reality conditions $\xi_n^* = \xi_{-n}$ and  $\psi_n^* = \psi_{-n}$) and the measure of the path integral (\ref{path}) is 

\begin{equation} \label{36}
{\cal D}\xi^\mu {\cal D} \psi ^\nu = d\xi_0 d\psi _0 \prod_{n =1}^{\infty}
d\psi _n d\psi _{-n} d\xi_n d\xi_{-n} ,
\end{equation}
where $d\xi_n$ is a abbreviation for $d\xi_n^1 \cdots d\xi_n^4$, etc.  

Then, the integrals over $\psi_n$ and $\xi_n$ ($n \neq 0$) can performed by standard techniques \cite{Windey,Chandia} and

\begin{eqnarray} \label{ind'}
\mbox{ind}\hat{Q} = {\cal N} & &\int  d\xi_0 d\psi_0 e^{\frac{1}{2}N_{\mu\nu\lambda\rho}\psi_0^\mu \psi_0^\nu \psi_0^\lambda \psi_0^\rho} \nonumber \\
& & \left[ \prod_{n=1}^{2} \frac{R_n}{\sinh (R_n)} +2 f_{\mu \nu} f_{\lambda \rho} \psi_0 ^{\mu}\psi_0 ^{\nu}\psi_0 ^{\lambda}\psi_0 ^{\rho}\right],
\end{eqnarray}
where $R_n$ are the components the antisymmetric matrix ${\bf R}$ in a canonical basis (generically ${\bf R}_{\mu\nu} = R_{\mu\nu\lambda\rho} \psi^\lambda \psi^\rho$),
 
\begin{equation} \label{R}
{\bf R} = \left( \begin{array}{cccc}  
0 & R_1 & 0 & 0 \\
-R_1 & 0 & 0 & 0  \\
0 & 0  &  0 & R_2 \\  
0 & 0  & -R_2 & 0 \\ \end{array} \right),
\end{equation}
$f_{\mu\nu}= \partial_{\mu} T^{\alpha}_{\alpha \nu} -  \partial_{\nu} T^{\alpha}_{\alpha \mu} $, and ${\cal N}$ is a normalization constant.

Since ${\bf R}$ is quadratic in the Grassmanian variables $\psi_0$, only terms proportional to $\psi_0^\mu \psi_0^\nu \psi_0^\lambda \psi_0^\rho$ survive in the expansion of the integrand in (\ref{path}). Upon integration over $d\psi_0$, we arrive to the index for the Dirac operator can be written in differential forms as

\begin{equation}
{\mbox {Ind}} D = \frac{1}{8\pi^2} \int [R^{ab} \mbox{\tiny $\wedge$} R_{ab} + 2 (T^a \mbox{\tiny $\wedge$} T_a - R_{ab} \mbox{\tiny $\wedge$} e^a \mbox{\tiny $\wedge$} e^b)],
\end{equation}
where the normalization constant ${\cal N}$ has been chosen equal to $\frac{1}{8\pi^2}$.  The $f^2$-term has been dropped out from the result, since the "Abelian $U(1)$ contribution" $\int f^2$ is zero because $\pi_3[U(1)] =0$. 

\section{Conclusion and Discussion}

We have obtained a torsional contribution to the index for the Dirac operator defined on a curved manifold. As the integral of the chiral anomaly is this index, the computation performed here is an independent derivation of the result obtained in \cite{CZ}. 

As the index for the Dirac operator with or without torsion is an integer, the integral of the Nieh-Yan form is necessarily an integer as well.  In fact, the integral of $N$ is the difference of the Chern classes for $SO(5)$ and $SO(4)$.  Since ${\pi_3} [SO(5)] = {\bf Z}$ and ${\pi_3} [SO(4)] = {\bf Z} + {\bf Z}$, one concludes that the Nieh-Yan can always be written as the sum of three integers. Field configurations with nontrivial Nieh-Yan numbers were constructed in Ref. \cite{CZ}.

The index theorem for a space with torsion had been computed previously in \cite{Mavromatos} finding no additional contribution to the torsion-free case.  That analysis precisely assumes $d[e_a\wedge T^a]\equiv 0$, so the result is hardly surprising from our point of view. 

The chiral anomaly corresponds to an observable effect, namely the opening of a decay channel that classically could not have been guessed.  Our result shows that there could be a similar effect due to the presence of topologically nontrivial torsional configurations in spacetime. This could be of relevance in the early universe, where the geometry could vary appreciably over small enough regions.  However, the possibility of direct evidence of such effect are clearly quite remote at present.  For a discusion for Dirac operator in the context of string theory, see \cite{Bellisai}.  Evidence of topological effects due to torsion are also discussed in \cite{Unzicker}.  

There are very few direct experimental tests that show the coupling between particles and the curvature of spacetime.  Even fewer experimental tests can be imagined for the detection of torsion.  The most direct effect on a spin 1/2 field could be through the coupling (\ref{D}) \cite{Obukhov} which is almost indistinguishable from an electromagnetic interaction with $A_{\mu}$ substituted by $T^{\alpha}_{\alpha \mu}$.  The analysis here shows that a spinning particle picks up an additional coupling of the form $N_{\mu\nu\lambda\rho} J^{\mu\nu} J^{\lambda\rho}$, where $N$ are the components of the Nieh-Yan tensor and $J$ are the generator of local Lorentz group. In the semiclassical limit for massless particles, these generators are $J^{ij} = \epsilon^{ijk} \sigma_k$ and $J^{0i} = p^i/p^0$ and the interaction takes the form $h N$, where $h$ is the helicity $\vec{p}\cdot \vec{\sigma}/p^0$ and $N \equiv \epsilon^{\mu\nu\lambda\rho} N_{\mu\nu\lambda\rho}$.  This means that if spacetime carries torsion, it would produce a splitting between positive and negative helicity states analogous to the the Stern-Gerlach effect.

\section*{Acknowledgements}

Many enlightening discussions with  M. Ba\~nados,  D. Bellisai, N. Brali\'c, A. Gomberoff, S. Hojman, C. Mart\'{\i}nez, F. M\'endez, J. Pereira, R. Troncoso are acknowledged. This work was supported in part by grants 1960229, 1970151 and 1980788 from FONDECYT (Chile), and 27-953/ZI and 27-953/GR from DICYT (USACH). Institutional support to CECS from Fuerza A\'erea de Chile and a group of Chilean private companies (Business Design Associates, CGE, CODELCO, COPEC, Empresas CMPC, Minera Collahuasi, Minera Escondida, NOVAGAS,  and XEROX-Chile) is also acknowledged.

\end{document}